\documentclass[aps,prl,amsmath,amssymb,twocolumn,preprintnumbers,nofootinbib]{revtex4}
\pdfoutput=1
\usepackage{graphicx}
\usepackage{multirow}
\usepackage{color}
\newcommand{\beq}{\begin{eqnarray}}
\newcommand{\eeq}{\end{eqnarray}}

\newcommand{\bmp}{\noindent\begin{minipage}{16cm}}
\newcommand{\emp}{\end{minipage}\vskip 7mm} 

\usepackage{dcolumn}
\usepackage{bm}
\usepackage{bbm}
\usepackage{pxfonts}
\usepackage{cancel} 
\usepackage{xspace}

\usepackage[margin=5pt, font=normalsize,labelfont=bf,format=plain,justification=centerlast]{caption}
\usepackage{subfig}

\definecolor{rossoCP3}{cmyk}{0,.88,.77,.40}

\def\lsim{\mathrel{\rlap{\lower4pt\hbox{\hskip1pt$\sim$}}
    \raise1pt\hbox{$<$}}}                
\def\gsim{\mathrel{\rlap{\lower4pt\hbox{\hskip1pt$\sim$}}
    \raise1pt\hbox{$>$}}}                

\newcommand{\be}{\begin{eqnarray}}
\newcommand{\ee}{\end{eqnarray}}

\newcommand{\ea}[1]{
\begin{align}
#1
\end{align}
}

\newcommand{\nonBZ}{{\cancel{BZ}}}

\begin{document}
\title{\Large  \color{rossoCP3} a New Conformal Window Bound from the a theorem}
\author{Oleg {\sc Antipin} }
\email{antipin@cp3.dias.sdu.dk} 
\author{Marc {\sc Gillioz} }
\email{gillioz@cp3.dias.sdu.dk} 
\author{Francesco {\sc Sannino}}
\email{sannino@cp3.dias.sdu.dk} 
\affiliation{\vskip .1cm\\
{ \color{rossoCP3}  \rm CP}$^{\color{rossoCP3} \bf 3}${\color{rossoCP3}\rm-Origins} \& the Danish Institute for Advanced Study {\color{rossoCP3} \rm DIAS},\\ 
University of Southern Denmark, Campusvej 55, DK-5230 Odense M, Denmark.}
\begin{abstract}
We propose a novel constraint on the gauge dynamics of strongly interacting gauge theories stemming from the a theorem. The inequality we suggest is used to provide a lower bound on the conformal window of four dimensional gauge theories.  \\[.1cm]
{\footnotesize   \it Preprint: CP$^3$-Origins-2012-4 \& DIAS-2012-4}
\end{abstract}

\maketitle

Gauge theories provide a remarkable description of physical phenomena till hundreds of GeVs. 
However our predictive power is severely hampered by our poor understanding of the strongly interacting regime of these theories. In fact, even for the time-honoured theory of nature, i.e. Quantum Chromodynamics (QCD) we are not yet able to compute all the physical quantities of interest, e.g., the full spectrum of the theory. 

To tame the nonperturbative regime of four dimensional gauge theories several tools have been proposed and used, from 't Hooft anomaly matching conditions~\cite{'tHooft:1982} to Cardy's conjecture  \cite{Cardy:1988cwa} of the existence of a four dimensional analog of Zamolodchikov $c$ theorem~\cite{Zamolodchikov:1986gt}, the $a$ theorem. The  $c$ theorem in two dimensions establishes the irreversibility of the renormalisation group (RG) flow and the $a$ theorem is thought to be its four dimensional generalisation. 

Here we propose an inequality limiting the loss of degrees of freedom of a theory when flowing from the ultraviolet  (UV) to the infrared (IR). We also show that the inequality provides relevant bounds for several four dimensional gauge theories, some of which are currently intensively studied via first principle lattice simulations. 

\section{The Inequality}

In four dimensions and for a general quantum field theory the vacuum expectation value of the trace of the energy-momentum tensor for a locally flat metric $g_{\mu \nu}$ reads
\begin{equation}
	\left\langle T_\mu^\mu \right\rangle = c \, W^2(g_{\mu\nu})
		- a \, E_4(g_{\mu\nu}) + \ldots \ .
\end{equation}
with $a$ and $c$ real coefficients, $E_4(g_{\mu\nu})$ the Euler density and $ W(g_{\mu\nu})$ the Weyl tensor. The dots represent contributions coming from operators that can be constructed out of the fields defining the theory. The contribution of these operators is proportional to the beta function of the associated couplings. The coefficient $a$ is the one used in Cardy's conjecture and for a free field theory reads~\cite{Duff:1977ay}
\begin{equation}
	a_{free} = \frac{1}{90 (8\pi)^2} \left( n_s + \frac{11}{2} n_f + 62 n_v \right),
	\label{eq:afree}
\end{equation}
where $n_s$, $n_f$ and $n_v$ are respectively the number of real scalars, Weyl fermions and gauge bosons. 

The change of $a$ along the RG flow is directly related to the underlying dynamics of the theory via its beta functions.  A generic renormalisable four dimensional gauge theory features several couplings. Here, we assume, for simplicity, that the theory possesses only one  gauge coupling $\alpha = g^2/4\pi^2$.  A more general analysis is presented in \cite{Antipin:2013xxx}. 
 Jack and Osborn~\cite{Osborn:1989td, Jack:1990eb} proposed that rather than utilising $a$ one should employ the function $\tilde{a}$ related to $ a$ via 
\begin{equation}
	\tilde{a} = a + W \beta_\alpha \ ,
	\label{eq:atilde}
\end{equation}
with $W$ another function of the coupling $\alpha$ and $\beta_{\alpha}$ the beta function of the coupling $\alpha$. At fixed points $a$ and $\tilde{a}$ coincide.
 
The Weyl consistency condition for  $\tilde{a}$ implies
\begin{equation}
	\partial_\alpha \tilde{a} = -\frac{n_v}{128\pi^2} \left( 1 + A \alpha + B \alpha^2 + \ldots \right)\beta_\alpha  \ ,
	\label{eq:consistencycondition}
\end{equation}
where $A$ and $B$ are group theoretical factors \cite{Jack:1990eb} which will not be relevant here. From this expression  one finds 
\begin{equation}
	\mu \frac{d\tilde{a}}{d\mu} = \beta_\alpha \partial_\alpha \tilde{a}
		=  -\frac{n_v}{128\pi^2} \left( 1 + A \alpha + B \alpha^2 + \ldots \right)\beta_\alpha^2  \leq 0 \ .
\label{dalpha}
\end{equation}
The last inequality holds at the leading order in perturbation theory and has been conjectured to be valid also beyond perturbation theory \cite{Komargodski:2011xv,Komargodski:2011vj}. This implies that the change in $a$ or $\tilde{a}$  between an UV fixed point and an IR one is positive.

Let's turn our attention to the investigation of the change in ${a}$ for an asymptotically free gauge theory featuring an IR fixed point. We will be evaluating the change in $a$ along the RG flow connecting the trivial UV fixed point to the IR one. 
Furthermore, we assume that by changing continuously one of the parameters of the theory, such as the number of flavours, the IR fixed point can be lost. We define $a_{lost}^{IR}$  as the value of $a$ at the fixed point just before large distance conformality is lost, and declare with $\Delta a = a_{free} - a_{lost}^{IR}$ the change in $a$ at this fixed point. Positivity of $a$  ensures
\begin{equation}
 a_{free}  \geq  \Delta a   \ .
 \label{AGS}
\end{equation}
One can imagine several mechanisms responsible for  the destruction of the IR fixed point \cite{Sannino:2012wy}. Here we postulate that it merges with an UV one. To elucidate this possibility we determine now $\Delta a$ assuming a merger phenomenon occurring in perturbation theory which has been actually observed in~\cite{Antipin:2013xxx}. The following $\beta$-function  encapsulates this phenomenon
\begin{equation}
	\beta_{\alpha} = -2\alpha^2 \left[b_0 + b_1 \alpha + b_2 \alpha^2 \right],
	\label{effbeta}
\end{equation}
with the two zeros at
\begin{eqnarray}
\label{merger}
\alpha^{BZ}& = &-\frac{b_1}{2b_2}\left(1 -  \sqrt{1-\frac{4b_0 b_2}{b_1^2}} \right) \ , \\ 
\alpha^\nonBZ& = &-\frac{b_1}{2b_2}\left(1+ \sqrt{1-\frac{4b_0 b_2}{b_1^2}}\right)   \ .
\end{eqnarray}
The negative sign solution corresponds, for small $b_0$, to the usual Banks-Zaks (BZ) fixed point. 
The two fixed points merge for $b_1^2 = 4 b_0 b_2$ and at the merger the coupling is $\alpha^{merger} = -2b_0/b_1$.  If we insist in using perturbation theory for both the IR and UV fixed points, then besides having a small $b_0$ one should also tune $b_1$ to be small and approaching zero as $\sqrt{b_0}$. This allows a consistent counting in perturbation theory as shown in \cite{Antipin:2013xxx}. The perturbative expression for $\Delta a$  reads
\begin{eqnarray}
	\Delta{a} 
	& = & \frac{n_v}{64\pi^2} \left[ b_0 \, \alpha^{merger} 
		+ \frac{1}{2} \left( b_1 + A b_0 \right) (\alpha^{merger})^2  \right. \nonumber \\
		&&\left. + 
		 \frac{1}{3} \left( b_2 + A b_1+ B b_0\right) (\alpha^{merger})^3 \right] \ ,
		\label{effa} 
\end{eqnarray}
with $A$ and $B$ the same group theoretical factors as in \eqref{dalpha} that can be neglected for small $b_1$ and $b_0$. Therefore in this limit we have
\ea{
 {\Delta{a}} = - \frac{n_v}{96\pi^2}   \frac{b_0^2}{b_1}   \ .
}
The expression above only depends on the two universal coefficients of the beta function. In addition it is not limited to pure gauge theories but can be used also for theories with Yukawa interactions  provided one replaces $b_1$ with the effective coefficient described in \cite{Antipin:2013xxx}.

To be predictive beyond perturbation theory we propose that the above expression holds also non perturbatively.  This allows us to bound the number of degrees of freedom, at the merger, by imposing the inequality \eqref{AGS} which now reads
\begin{equation}
 n_s + \frac{11}{2} n_f + 62 n_v  \geq  60 \, n_v  \frac{b_0^2}{|b_1|}  \ .
 \label{AGS-practical}
\end{equation}
As we shall see the inequality above provides strong constraints on the number of degrees of freedom at the merger.  If the merger bounds the region of conformality then we are able to provide a lower bound on the size of the conformal window for generic gauge theories.  
 \section{Conformal Window estimates}
The conformal window is defined as the region in theory space, as function of number of flavours and colours, where the underlying asymptotically free gauge theory displays large distance conformality.  The upper boundary of the conformal window is determined by setting to zero the first coefficient of the beta function $b_0$, while to determine the lower boundary for nonsupersymmetric gauge theories  we use here the inequality \eqref{AGS-practical}. 

\subsection{$SU(N_c)$ Gauge Theories}
 We start with gauge theories with $N_f$ Dirac fermions in a given representation $R$ of an $SU(N_c)$ gauge group, for which $b_0= \frac{11}{3}N_c - \frac{4}{3}N_f T(R)$, and $3b_1 =  34 N_c^2 - 20 N_c T(R) N_f - 12 C_2(R)T(R)N_f $. $T(R)$ is the trace normalisation of the gauge generators in the representation $R$, and $C_2(R)$ is the associated quadratic Casimir. For these theories the bound \eqref{AGS-practical} reads
\begin{eqnarray}
{11} \frac{ N_f d(R)}{N_c^2 - 1} + 62 \geq  \frac{20 \, (11N_c - 4 N_f T(R))^2}{\Big| 34 N_c^2 - 20 N_c T(R) N_f - 12 C_2(R)T(R)N_f \Big|} \ . \nonumber \\
\end{eqnarray}
Here $d(R)$ is the dimension of the representation. For the fundamental representation at large $N_c$ the inequality requires   $N_f \geq 3.01 N_c$. For $N_c=3$ we find $N_f \geq 9.18$, while for $N_c=2$ we have $N_f \geq 6.24$. 

For the two-index symmetric representation we find for $N_c=2$ and $N_c=3$ respectively $N_f\geq 1.33$ and $N_f \geq 1.55$. 
These results are surprisingly close to different estimates of the conformal window summarised in \cite{Sannino:2009za}, while being the first estimates making direct use of the $a$ theorem. 

\subsection{Goldstone Bosons Corrections}
Below the lower boundary of the conformal window, and for QCD-like theories, we expect the quantum global symmetry of the theory to break spontaneously according to the following patterns: For $N_f$ Dirac fermions in the complex representation of the gauge group, the global symmetry $SU(N_f)\times SU(N_f)\times U(1)$ is expected to break to $SU(N_f)\times U(1)$ yielding $N^2_f -1$ Goldstone bosons; For real representations (like the adjoint) the symmetry group is enhanced to $SU(2 N_f)$ and it is expected to break spontaneously to $SO(2N_f)$, yielding $(N_f^2 + N_f -2)/2$ Goldstone bosons; Finally for pseudoreal representations (e.g. two colour QCD) the global symmetry group $SU(2 N_f)$  is expected to break to $Sp(2 N_f)$. We note that these last two, less familiar patterns of chiral symmetry breaking have been extensively reviewed in \cite{Sannino:2009aw} and have been recently shown to occur for two phenomenologically relevant examples respectively in \cite{Hietanen:2012sz} and \cite{Lewis:2011zb} using first principle lattice simulations.  

Taking into account the fact that Goldstone bosons interact only derivatively at low energy, the inequality \eqref{AGS-practical} gets corrected as follows
\begin{equation}
 n_s - n_{GB} + \frac{11}{2} n_f + 62 n_v  \geq  60 \, n_v  \frac{b_0^2}{|b_1|}  \ ,
 \label{AGS-GB}
\end{equation}
where $n_{GB}$ is the number of real Goldstone bosons. Furthermore it becomes clear that for theories without a mass gap below the conformal window the bound in \eqref{AGS-practical} cannot be saturated. We checked that the effect of the Goldstone bosons on the size of the conformal window is tiny and practically irrelevant for the higher dimensional representations. Incidentally, this is the reason why Cardy's bound, in the absence of the contribution on the right-hand-side of \eqref{AGS-GB}, is not sufficiently constraining.  

We report here for completeness only the values of the critical number of flavours for the fundamental representation. At large $N_c$ we have $N_f \geq 3.04 N_c$, for $N_c=3$ the critical value is $N_f \geq 9.28$, and finally for $N_c =2$ we have $N_f \geq 6.38$.

The results presented in this work must be taken {\it cum grano salis} because of our use of \eqref{AGS-practical} and \eqref{AGS-GB}  beyond perturbation theory. Nevertheless we feel that the results are encouraging and elucidate the potential predictive power of the $a$ theorem.

\subsection{Comparison with Supersymmetry}

Although for supersymmetric gauge theories one does not expect the IR fixed point to annihilate with an UV one, it is instructive to determine the lower boundary of the super QCD conformal window using the inequality \eqref{AGS-practical}. For this theory we have
\begin{equation}
\frac{ 15 N_f N_c }{N_c^2 - 1} + \frac{11}{2}+  62 \geq  60 \frac{ \, (3N_c -  N_f)^2}{\Big| 6 N_c^2 - 2N_f(2N_c  - \frac{1}{N_c}) \Big|} \ .
\end{equation}
At large $N_c$ one finds $N_f \geq 7/4 = 1.75 N_c$ which is higher than Seiberg's result of $1.5 N_c$ \cite{Seiberg:1994pq}. The reason why our result leads to a smaller conformal window  is because $b_1$ appears in the denominator on the right-hand-side of \eqref{AGS-practical} and it vanishes for a value $N_f$ higher than the critical value found by Seiberg, as observed in~\cite{Ryttov:2007sr}.

\section{Conclusions}
We have proposed a new bound on the conformal window of strongly interacting nonsupersymmetric gauge theories. We used an inequality rooted  in the $a$ theorem.  The assumption that the lower boundary of the conformal window coincides with the merging of two nontrivial fixed points has also been employed.  We first computed the change in $a$ at the merger and then used it to estimate the lower boundary of the conformal window. The results are in agreement with earlier estimates and show how the $a$ theorem can be of use. It would be interesting to test the bound for other nonsupersymmetric gauge theories, and, when possible, also use first principle lattice simulations to determine the actual change in $a$ in the nonperturbative regime and compare it with the estimates presented here. 

\vskip .3cm
The CP$^3$-Origins centre is partially funded by the Danish National Research Foundation, grant number DNRF90. 

\bibliography{biblio}

\end{document}